\documentclass[a4paper,12pt]{article}
\usepackage{epsfig}
\usepackage{citesort}
\date{\today}
 \normalsize

\newcommand{\bmat}{\left(\begin{array}}
\newcommand{\emat}{\end{array}\right)}
\newcommand{\be}{\begin{equation}}
\newcommand{\ee}{\end{equation}}
\newcommand{\bea}{\begin{eqnarray}}
\newcommand{\eea}{\end{eqnarray}}

\newcommand{\W}{{\scriptscriptstyle W}}

\newcommand{\alfasw}{\alpha_s({m_{\W}})}
\def\du#1#2{{\left(\delta^u_{#1}\right)_{#2}}}
\def\dd#1#2{{\left(\delta^d_{#1}\right)_{#2}}}

\def\lsim{\raise0.3ex\hbox{$\;<$\kern-0.75em\raise-1.1ex\hbox{$\sim\;$}}}
\def\gsim{\raise0.3ex\hbox{$\;>$\kern-0.75em\raise-1.1ex\hbox{$\sim\;$}}}
\def\Frac#1#2{\frac{\displaystyle{#1}}{\displaystyle{#2}}}

\def\susy{\mbox{\tiny SUSY}}
\def\sm{\mbox{\tiny SM}}

\begin{document}
\renewcommand{\thefootnote}{\fnsymbol{footnote}}
\rightline{IPPP/03/30} \rightline{DCPT/03/60}
\rightline{HIP-2003-33/TH}
\vspace{.3cm} 
{\Large
\begin{center}
{\bf Chargino contributions to the CP asymmetry  in $B\to \phi K_S$ decay}
\end{center}}
\vspace{.3cm}
\begin{center}
D. Chakraverty$^{1}$, E. Gabrielli$^{1}$, K. Huitu$^{1,2}$, 
and S. Khalil$^{3,4}$\\
\vspace{.3cm}
$^1$\emph{Helsinki Institute of Physics, POB 64, 00014 University of
Helsinki, Finland.}\\
$^2$\emph{Div. of HEP, Dept. of Phys. Sciences, POB 64,00014
University of Helsinki, Finland.}\\
$^3$ \emph{IPPP, University of Durham, South Rd., Durham
DH1 3LE, U.K.}\\
$^4$ \emph{Ain Shams University, Faculty of Science, Cairo, 11566, Egypt.}\\
\end{center}

\vspace{.3cm}
\hrule \vskip 0.3cm
\begin{center}
\small{\bf Abstract}\\[3mm]
\end{center}
We perform a model independent analysis of 
the chargino contributions to the CP asymmetry in 
$B\to \phi K_S$ process. We use the mass insertion approximation 
method generalized by including the possibility of a light right-stop.  
We find that the dominant effect is given by 
the contributions of the mass insertions $(\delta_{LL}^u)_{32}$ 
and $(\delta_{RL}^u)_{32}$  to the Wilson coefficient of the  
chromomagnetic operator. 
By considering both these contributions simultaneously, 
the CP asymmetry in $B\to \phi K_S$ process
is significantly reduced and negative values, which are
within the 1$\sigma$ experimental range and satisfy the 
$b\to s \gamma$ constraints, can be obtained.
\begin{minipage}[h]{14.0cm}
\end{minipage}
\vskip 0.3cm \hrule \vskip 0.5cm
The measurement of CP asymmetries in nonleptonic B decays
plays a crucial role in testing the CP violation mechanism 
of the Standard Model (SM) and it is a powerful probe of New Physics 
(NP) beyond the SM.
The CP asymmetries are usually described by the time dependent rates 
$a_{f_{CP}}(t)$, for $B^0$ and $\bar{B}^0$ to a CP eigenstate $f_{CP}$ 
\begin{eqnarray}
a_{f_{CP}}(t)&=&\frac{\Gamma (\overline{B}^0(t)\to f_{CP})-\Gamma
(B^0(t)\to f_{CP})} {\Gamma (\overline{B}^0(t)\to f_{CP})+\Gamma (B^0(t)
\to f_{CP})} 
\nonumber \\
&=&C_{f_{CP}}\cos\Delta M_{B_d}t+S_{f_{CP}}\sin\Delta M_{B_d}t
\label{CPasym}
\end{eqnarray}
where $C_{f_{CP}}$ and $S_{f_{CP}}$ represent the coefficients 
of direct  and indirect CP violations respectively, and
$\Delta M_{B_d}$ is the $B^0$ eigenstate mass difference.

The time dependent CP asymmetry 
$a_{J/\psi K_S}(t)$ in the B meson decay 
$B\to J/\Psi K_S$ has been recently measured 
by BaBar and Belle Collaboration, with an average of 
$S_{J/\Psi K_S} = \sin{2\beta}=0.734 \pm 0.034$ \cite{babar,belle},
showing the first evidence of CP violation in B meson system 
in perfect agreement with the Standard Model (SM) predictions.
This is expected, since the SM contribution is at tree-level.

For the decay $B\to \phi K_S$, where the same weak phase is measured,
the situation is qualitatively different.
The SM contribution is at one-loop level, and one can expect crucial
contributions from New Physics.
The branching ratio for $B\to \phi K_S$ has recently 
been measured by both BaBar and Belle \cite{Br}
with an average for the branching ratio of $BR(B\to \Phi K_S) =
\left(8.4^{+2.5}_{-2.1}\right)\times 10^{-6}$ which is
slightly different from the SM predictions. However, this is not 
a signal of a real problem, since the SM evaluation of $BR(B\to \Phi K_S)$
is largely affected by theoretical uncertainties in the 
evaluation of hadronic matrix elements.
On the other hand, the time dependent CP asymmetry in Eq.(\ref{CPasym})
is less sensitive to these uncertainties, since the hadronic 
matrix elements almost cancel out in the ratio of rates.

Recently BaBar and Belle Collaborations 
\cite{babar2,belle} have also measured
the time dependent CP asymmetry in $B\to \phi K_S$ process, reporting 
an average value of $S_{\phi K_S} = -0.39 \pm 0.41$. 
In the SM, $S_{\phi K_S}$ is 
expected to give the same value of $\sin 2\beta$ as extracted from 
$S_{J/\psi K_S}$, up to terms of order $O(\lambda^2)$
where $\lambda$ is the Cabibbo mixing. Thus, the comparison of
the experimental results for  $S_{J/\psi K_S}$ and $S_{\phi K_S}$ 
reveals a $2.7~ \sigma$ deviation from the SM prediction. 
If this discrepancy will be confirmed with a better accuracy, it will be
a clean signal of NP.

Due to the 
additional sources of flavor and CP violation beyond the ones of 
Cabibbo-Kobayashi-Maskawa (CKM) mixing matrix,
supersymmetric (SUSY) models are natural candidates
for explaining the difference between  the CP asymmetries $S_{\phi K_S}$ 
and $S_{J/\psi K_S}$. 
Recently, the gluino contributions to 
$S_{\phi K_S}$ have been analyzed in Refs.\cite{Khalil:2002fm,gluino}. 
In these works, it has been shown 
that gluino exchanges can explain the experimental 
results of $S_{\phi K_S}$ without conflicting the 
experimental constraints from $S_{J/\psi K_S}$ and the branching ratio 
$BR(b\to s \gamma)$.

The main purpose of this letter is to show that also the chargino 
contributions to $S_{\phi K_S}$ can be significant  
and account for these recent measurements. 
We perform a model independent analysis by using the well known
method of mass insertion approximation \cite{Hall:1985dx}, 
generalized by including the possibility of a light right-stop 
in the otherwise almost degenerate squark spectrum.
In our analysis, we take into account all the operators that contribute to the 
effective Hamiltonian for $\Delta B=1$ transitions 
$H_{\mathrm{eff}}^{\Delta B=1}$ and provide analytical results 
for the corresponding leading Wilson coefficients.

Now we start our analysis of the SUSY contributions to the 
time dependent CP asymmetry in $B\to \phi K_S$ decay.
In the following we will adopt the parameterization of
the SM and SUSY amplitudes as in Ref.\cite{Khalil:2002fm}, namely
\be
\left(\frac{A^{\susy}}{A^{\sm}}\right)_{\phi K_S}\equiv 
R_{\phi}~ e^{i\theta_{\phi}}~e^{i\delta_{12}}
\label{ratio}
\ee
where $\theta_{\phi}$ is the SUSY CP violating phase,
and $\delta_{12}=\delta_{SM}-\delta_{SUSY}$
is the strong (CP conserving) phase.
In this case, the mixing CP asymmetry  
$S_{\phi K_S}$ takes the following form 
\begin{eqnarray}
S_{\phi K_S}\!&=&\!\Frac{\sin 2 \beta \!+\!2 R_{\phi} 
\cos \delta_{12} \sin(\theta_{\phi}\!+\!2 \beta)\!+
R_{\phi}^2 \sin (2 \theta_{\phi}\!+\!2 \beta)}{1+ 2 R_{\phi} 
\cos \delta_{12} \cos\theta_{\phi} +R_{\phi}^2}.
\label{cpmixing}
\end{eqnarray}
The most general amplitude for $B\to \phi K_S$ process can be written as 
\begin{equation}
\overline{A}(\phi K)=-\frac{G_F}{\sqrt{2}}\sum_{i=1}^{12}
\left[C_i(\mu)+\tilde{C}_i(\mu)\right]
\langle\phi\bar{K}^0 | Q_i (\mu)|\bar{B}^0\rangle ,
\end{equation}
where $Q_i$ are the operators which contribute to the
effective Hamiltonian for $\Delta B=1$ transitions and $C_i(\mu)$ are 
the corresponding Wilson coefficients at energy scale $\mu$.
The matrix elements $\langle\phi\bar{K}^0 | Q_i |\bar{B}^0\rangle$ are 
calculated in the naive factorization approximation \cite{Ali:1998eb},  
and their expressions can be found in Ref.\cite{Khalil:2002fm}. 
In this notation, the $Q_{i=1-10}$ represent the
four-fermion operators, and $Q_{11}$ and $Q_{12}$ the magnetic and 
chromomagnetic dipole operators respectively.
The Wilson coefficients $\tilde{C}_i$ are associated to the
operators $\tilde{Q}_i$ which are obtained from $Q_i$ by exchanging 
$\gamma_5\to -\gamma_5$ in their chiral structure, 
see Ref.\cite{Khalil:2002fm} for their definition. 
In the SM, $\tilde{C}_i$ are chirally 
suppressed with respect to $C_i$ ones by terms proportional to light quark
masses. However, in non-minimal 
SUSY extensions of the SM
they can receive sizeable contributions, for instance from the 
gluino mediated penguin and box diagrams. 
On the other hand, the 
chargino contributions to $\tilde{C}_i$ are always 
suppressed by Yukawas of the first two generations.
Thus, we can 
safely neglect $\tilde{C}_i$ contributions in our analysis.

The Wilson coefficients $C_i(\mu)$ at a 
lower scale $\mu \simeq \mathcal{O}(m_b)$ can be extrapolated by
the corresponding ones at high scale $C_i(\mu_W)$ as 
$ C_i(\mu) = \sum_j \hat{U}_{ij}(\mu, \mu_W) \, C_j(m_W)$, 
where $\hat{U}_{ij}(\mu, \mu_W)$ is the QCD evolution matrix and 
$\mu_W \simeq m_W$.
Since the operator $Q_{12}$ is of order $\alpha_s$, 
we include in our analysis the LO corrections only for the effective 
Wilson coefficient $C_{12}(\mu)$, 
while for the remaining ones $C_{i=1-10}(\mu)$
we use the matrix $\hat{U}_{ij}(\mu, \mu_W)$ at NLO order in QCD and QED
\cite{Buchalla:1995vs}. 

As is well known, supersymmetry affects the Wilson 
coefficients $C_i(\mu)$ only at high scale $\mu\simeq \mu_W$.
The chargino contributions to $C_i(\mu_W)$, corresponding to 
the effective Hamiltonian for $\Delta B=1$ transitions,
have been calculated exactly (at 1-loop) in Refs.\cite{BBMR} and \cite{GG}. 
Here we provide the results for chargino contributions to $C_i(\mu_W)$, 
evaluated at the first order in mass insertion approximation. 
By using the notation of Ref.\cite{GG} we obtain
\begin{eqnarray}
F_{\chi}\!&=\!&\Big[ \sum_{a,b} K^{\star}_{a2} K_{b3}\du{LL}{ba} \Big] 
R_F^{LL}
+ \Big[\sum_a K^{\star}_{a2} K_{33} \du{RL}{3a}\Big] Y_t \, R_F^{RL}
\nonumber \\
&+ &\Big[ \sum_a K^{\star}_{32} K_{a3} \du{LR}{a3} \Big] Y_t\, R_F^{LR}
+ \Big[ K^{\star}_{32} K_{33}\du{RR}{33} \Big] Y_t^2 R_F^{RR}
\label{Fch}
\end{eqnarray}
where for the definition of mass insertions $\du{AB}{ij}$ see  Ref.\cite{Hall:1985dx}.
Same notation of 
Ref.\cite{GG} has been used to relate $F$ quantities to 
the Wilson coefficients $C_{i=1-10}(\mu_W)$, while for the magnetic and 
chromomagnetic contributions we have $C_{11}(\mu_W)=M^{\gamma}$ and 
$C_{12}(\mu_W)=M^{g}$. 
Here $Y_t$ is Yukawa coupling of the top quark and 
$F$ refers to the photon-penguins ($D$),
$Z$-penguins ($C$), gluon-penguins ($E$), boxes with external down
quarks ($B^{(d)}$) and up quarks ($B^{(u)}$),
magnetic-penguins ($M^{\gamma}$), and 
chromomagnetic ($M^{g}$) penguin diagrams. 
We want to stress that there are also contributions from box diagrams 
mediated by both gluino and chargino exchanges, which affect only 
$C_{i=1,2}(\mu_W)$, but their effect is negligible \cite{GG}, \cite{new}
and we will not include them in our analysis. 

The detailed expressions for $R_{F}$, including 
contributions from chargino-gluino box diagrams, 
are given in the appendix.
Here we will just concentrate on the dominant contributions which turn out 
to be due to the chromomagnetic ($M^{g}$) penguin and 
$Z$-penguin ($C$) diagrams. In fact, 
for light SUSY particles ($\lsim 1$ TeV), 
the contribution from the chromomagnetic penguin is one order and two orders
of magnitudes larger than the corresponding ones from $Z$-penguin 
and other diagrams, respectively.
However, in our numerical analysis we take into 
account all the contributions.

{}From Eq. (\ref{Fch}), it is clear that $LR$ and $RR$ contributions are 
suppressed by order $\lambda^2$ or $\lambda^3$. Since we will work in 
$\mathcal{O}(\lambda)$ order, we can neglect them and simplify 
$F_{\chi}$ as, 
\begin{eqnarray}
F_{\chi} =  \Big[\;\du{LL}{32} + \lambda \du{LL}{31}\;\Big] R_F^{LL} 
+ \Big[\;\du{RL}{32} + \lambda \du{RL}{31}\;\Big]\, Y_t\, R_F^{RL}.
\label{Fchapprox}
\end{eqnarray}
The functions $R^{LL}_F$ and $R^{RL}_F$ depend on
the SUSY parameters through the chargino masses ($m_{\chi_i}$), 
squark masses ($\tilde{m}$) and the entries of
the chargino mass matrix. For instance for $Z$ and 
magnetic (chromomagnetic) dipole penguins $R_C^{LL,RL}$
and $R_{M^{\gamma (g)}}^{LL,RL}$ respectively, we have 
\begin{eqnarray}
R_C^{LL}&=&
\sum_{i=1,2}|V_{i1}|^2 \, P_C^{(0)}(\bar x_i)
+\sum_{i,j=1,2} \left[ U_{i1}V_{i1}U_{j1}^{\star}V_{j1}^{\star}\, 
P_C^{(2)}(x_i,x_j) \right. 
\nonumber\\
&+& \left. |V_{i1}|^2 |V_{j1}|^2
\left(\frac{1}{8}-P_C^{(1)}(x_i, x_j)\right)\right]
\nonumber
\\
R_C^{RL}&=&-\frac{1}{2}
\sum_{i=1,2}\, V_{i2}^{\star}V_{i1}\,  P_C^{(0)}(\bar x_i)
- \sum_{i,j=1,2}\, V_{j2}^{\star}V_{i1}\left(
U_{i1}U_{j1}^{\star}\, P_C^{(2)}(x_i, x_j)
\right.\nonumber \\
&+&
\left. 
V_{i1}^{\star} V_{j1}\, P_C^{(1)}(x_i, x_j)\right)
\nonumber
\\
R_{M^{\gamma, g}}^{LL}&=&\sum_i |V_{i1}|^2\,
x_{Wi}\, P_{M^{\gamma,g}}^{LL}(x_i) - Y_b\sum_i V_{i1} U_{i2}\,
x_{Wi}\, \frac{m_{\chi_i}}{m_b} P_{M_{\gamma,g}}^{LR}(x_i)
\nonumber
\\
R_{M^{\gamma, g}}^{RL}&=&
-\sum_i V_{i1}V_{i2}^{\star}\,
x_{Wi}\, P_{M_{\gamma,g}}^{LL}(x_i),
\label{Rterms}
\end{eqnarray}
where $Y_b$ is the Yukawa coupling of bottom quark, 
$x_{\W i}=m_W^2/m_{\chi_i}^2$, 
$x_{i}=m_{\chi_i}^2/\tilde{m}^2$, $\bar x_i =\tilde{m}^2/m_{\chi_i}^2$, and
$x_{ij}=m_{\chi_i}^2/m_{\chi_j}^2$.
The loop functions $P_C^{(1,2)}$, 
$P_{M_{\gamma, g}}^{LL(LR)}$ are given by
\bea
P_C^{(1,2)}(x,y)&=&-2 \left(x\frac{\partial }{\partial x}+
  y\frac{\partial }{\partial y}\right) C_{\chi}^{(1,2)}(x,y),~~
\nonumber \\
P_{M_{\gamma}}^{LL(LR)}(x)&=&- x \frac{d}{d x} \Big( 
x F_{1(3)}(x)+\frac{2}{3}x F_{2(4)}(x)\Big),~~~~
P_{M_{g}}^{LL(LR)}=-x \frac{d}{dx} \left(x F_{2(4)}(x) \right),
\eea
where $P_C^{(0)}(x)=-\lim_{y\to x} P_C^{(1)}(x,y)$,
and the functions $C_{\chi}^{(1,2)}(x,y)$ and  
$F_i(x)$ can be found in Refs.\cite{BBMR} and  \cite{GG}, respectively.
Finally, $U$ and $V$ are the matrices that diagonalize chargino mass matrix, 
defined as $U^* M_{\tilde{\chi}^+} V^{-1} = \mathrm{diag}(m_{\tilde{\chi}_1^+},
m_{\tilde{\chi}_2^+})$
where we adopted the notation of Ref.\cite{GG} for the chargino matrix 
$M_{\tilde{\chi}^+}$.

Notice that the dependence from Yukawa bottom $Y_b$ in Eq.(\ref{Fch})
leads to enhancing $C_{12}$ at large $\tan\beta$. 
Here, we also considered the case in which 
the mass of stop-right ($m_{{\tilde t}_R}$)
is lighter than other squarks. In this case
the functional form of Eq.(\ref{Fch}) remains unchanged, while
only the expressions of $R_F^{RL}$ should be modified by replacing
the functions inside $P_{M^{\gamma, g}}^{LL,RL}$
as $-x_i \frac{d}{d x_i} x_i F_a(x_i) \to {1\over{(x_t -1)}} 
\left[x_{it} F_a(x_{it}) -x_{i} F_a(x_{i})\right]$, with index $a=1-4$, 
$ P_C^{(1,2)}(x_i, x_j)\to {2\over{(x_t -1)}}
\left[ C_{\chi}^{(1,2)}(x_{jt},x_{it}) - C_{\chi}^{(1,2)}(x_{j},x_{i})\right]$,
and $P_C^{(0)}({\bar x}_i,{\bar x}_{it})={4\over{(x_t -1)}}
  \left[ C_{\chi}^{(1)}({\bar x}_{it},{\bar x}_{i}) - 
C_{\chi}^{(1)}({\bar x}_{i},{\bar x}_{i}) \right]$, 
where $x_{it}=\frac{m_{\chi_i}^2}{{m^2_{{\tilde t}_R}}}=1/\bar{x}_{it}$ and 
$x_t=\frac{{m^2_{{\tilde t}_R}}}{\tilde{m}^2}$.

In order to simplify our analysis,
we consider first the case where a mass insertion is dominant 
over the others. In this case we retain 
only the effect of a mass insertion per time, 
switching off all the others. Thus, there is only one 
SUSY CP phase which factorizes in the SUSY amplitude, and so 
$\theta_\phi$ in Eq. (\ref{ratio}) can be identified with 
the corresponding $arg[\du{AB}{ij}]$.

\begin{figure}[th]
\begin{center}
\hspace*{-7mm}
\epsfig{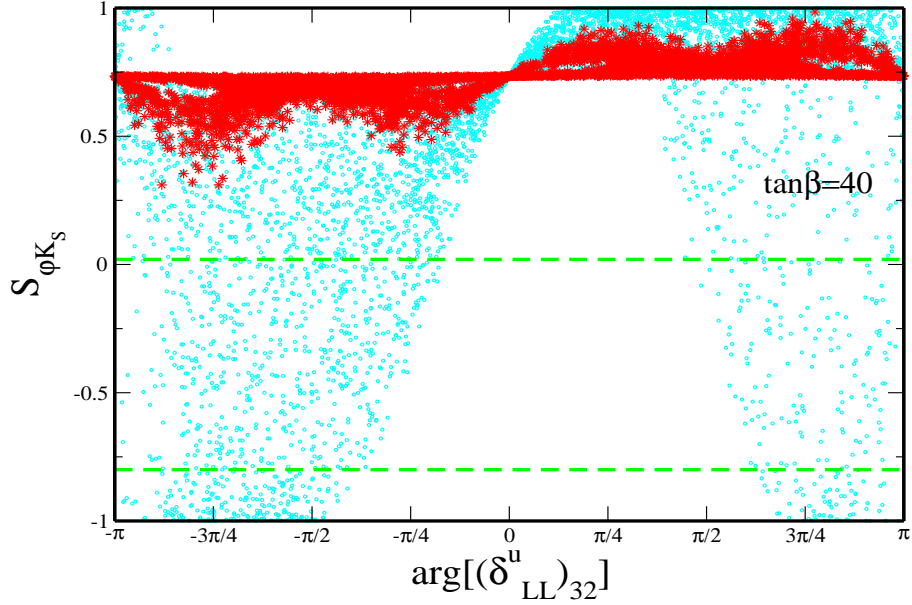}\\
\caption{The mixing CP asymmetry as function of 
$\arg[\du{LL}{32}]$, for $\tan \beta=40$, and with the contribution 
of one mass insertion $|\du{LL}{32}|$.
Red points correspond to $|\du{LL}{32}|$ that satisfy all
the experimental bounds.
The light blue points are not allowed by $BR(b \to s \gamma)$. 
The strong phase $\delta_{12}$ is fixed at $\cos\delta_{12}=1$.
}
\label{fig1}
\end{center}
\end{figure}

\begin{figure}[th]
\begin{center}
\hspace*{-7mm}
\epsfig{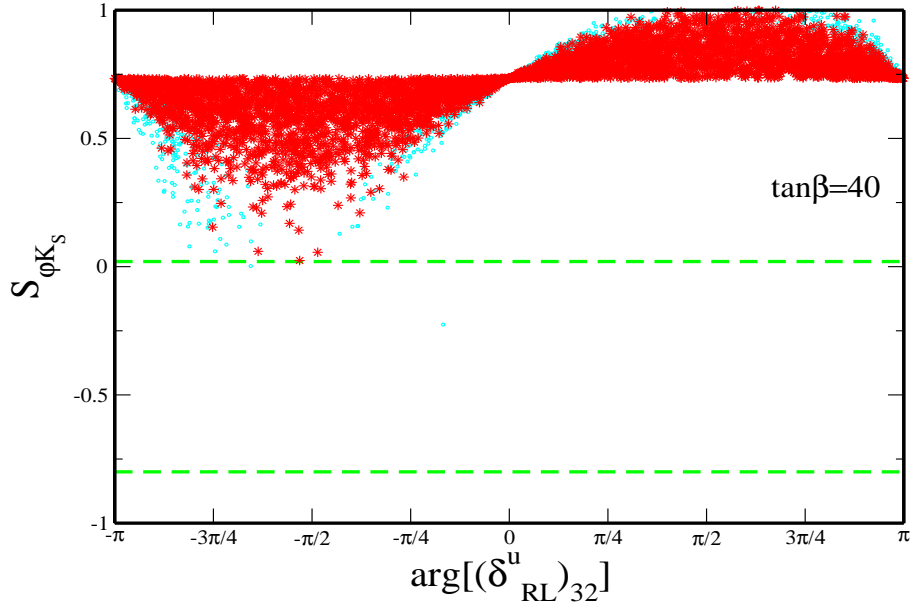}\\
\caption{As in Fig. \ref{fig1}, but for the mass insertion
$\du{RL}{32}$.}
\label{fig2}
\end{center}
\end{figure}

\begin{figure}[th]
\begin{center}
\hspace*{-7mm}
\epsfig{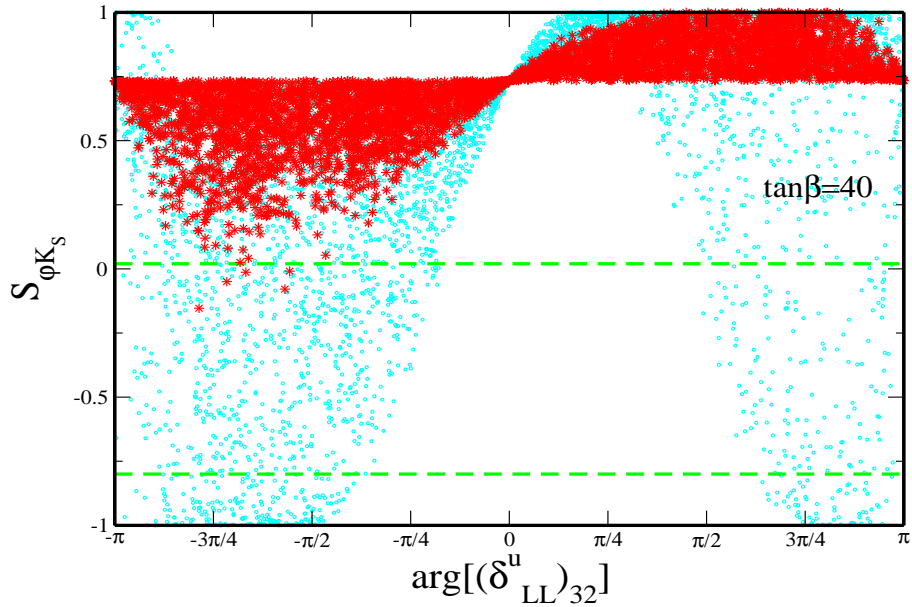}\\
\caption{The mixing CP asymmetry as function of 
$\arg[\du{LL}{32}]=\arg[\du{RL}{32}]$, for
$\tan \beta=40$, and with the contribution of two
mass insertions $|\du{RL}{32}|$ and $|\du{LL}{32}|$.
Red points correspond to $|(\delta_{LL}^d)_{32}|$ and
$|(\delta_{RL}^d)_{32}|$ that satisfy all the experimental 
bounds.
The light blue points are not allowed by $BR(b \to s \gamma)$. 
The strong phase $\delta_{12}$ is fixed at $\cos\delta_{12}=1$.}
\label{fig3}
\end{center}
\end{figure}

We present our numerical 
results in Figs. \ref{fig1}-\ref{fig3}, where 
the CP asymmetry $S_{\Phi K_S}$ is plotted versus the SUSY CP violating phase.
In this analysis we worked at fixed values of $\tan{\beta}$ 
and scanned over all the relevant SUSY parameters -
$\tilde{m}$, the weak gaugino mass $M_2$, the $\mu$ term, 
and $m_{{\tilde t}_R}$ - and required that they
satisfy the present experimental lower mass bounds, namely
the lightest chargino $m_{\chi} > 90$ GeV, heavy squarks
$\tilde{m} > 300$ GeV, and light right-stop $m_{{\tilde t}_R} > 150$ GeV.
In addition, we scanned over the real and imaginary 
part of the corresponding mass insertions, by requiring that 
the $b\to s \gamma$ and $B-\bar{B}$ mixing constraints
are satisfied. In our calculation we have used the 
formula of the  branching ratio (BR) $b\to s \gamma$ at the NLO in QCD, as 
provided in Ref.\cite{Kagan:1998bh}. Indeed, the BR of  $b\to s \gamma$
can be easily parametrized in terms of the SUSY contributions to 
Wilson coefficients $C_{11}$ and $C_{12}$ at $\mu_W$ scale given in 
Eq. (\ref{Fch}). For this parametrization, we used the central values of the 
SM parameters as provided in Ref.\cite{Kagan:1998bh}, and the low energy 
renormalization scale fixed at $\mu=m_b$.

In Figs.\ref{fig1} and \ref{fig2} we show 
the effects of one mass insertion 
per time, $\du{LL}{32}$ and  $\du{RL}{32}$, evaluated at
$\tan{\beta}=40$. In all these plots, the red points are allowed by
all experimental constraints, while
light-blue points correspond to the
points disallowed by $BR(b\to s \gamma)$ constraints 
at 95\% C.L. , namely
$2.0\times 10^{-4} < BR(b\to s\gamma) <  4.5\times 10^{-4}$.
In order to get the maximum effect for the
negative values of CP asymmetry, we fixed 
the strong CP conserving phase $\delta_{12}$ to be zero.
We have not shown the contributions of the other mass 
insertions since they are sub-leading, being 
suppressed by terms of order $\lambda$.

As we can see from the results in Figs.\ref{fig1}-\ref{fig2}, 
there is no chance with only one mass insertion 
to achieve negative values for the CP asymmetry. 
The main reason for $\du{LL}{32}$ is due to the $b\to s \gamma$
constraints which are particularly sensitive to $\tan{\beta}$, while 
this is not the case for $\du{RL}{32}$. Clearly, we have 
considered also different values of $\tan{\beta}$, and we found that
the allowed regions in the scatter plots are not very sensitive to 
$\tan{\beta}$.

In Fig. \ref{fig3} we show another example, where we take simultaneously 
both the mass insertions $\du{LL}{32}$ and $\du{RL}{32}$
per time, but assuming that their CP violating phase is the same.
As can be seen from Fig. \ref{fig3} there are points, allowed by 
$b\to s \gamma$ constraints,  which can fit inside the $1\sigma$ experimental 
region. 

In order to understand the behavior of these results, it is very useful 
to look at the numerical parametrization of the ratios 
of amplitudes in terms of the relevant mass insertions.
Indeed, we would like to show that the main contribution to the
SUSY amplitude is provided by the chromomagnetic dipole operator.
For example, with $M_2=200$ GeV, 
$\mu=300$ GeV, $m_{\tilde{q}}=400$ GeV, $m_{\tilde{t}_R}=150$ GeV, and
$\tan \beta=30$, we find 
$R^{RL}_C\simeq -0.033$, $R^{LL}_{M^g}\simeq -0.068$, while 
for all the other ones $R^{AB}_F \simeq O(10^{-3})$, 
and the amplitudes ratio 
$R_A\equiv \frac{A^{SUSY}}{A^{SM}}$  is given by
\be
R_A \simeq
0.37 \du{LL}{31} + 1.64 \du{LL}{32} - 0.05 
\du{RL}{31}  -0.21 \du{RL}{32}.
\ee
Now, if we switch off the chromomagnetic dipole operator,
the coefficients of the mass insertions
$\delta^u_{LL}$ are significantly reduced, while the coefficients 
of $\delta^u_{RL}$ are slightly changed 
and $R_A$ takes the form
\be
R_A \simeq 
-0.0031 \du{LL}{31} - 0.014 \du{LL}{32} - 0.045 \du{RL}{31} - 0.20 \du{RL}{32}.
\ee
It is worth mentioning that the chromomagnetic contributions are 
sensitive to the value of $\tan \beta$. Indeed, the contribution 
coming from $R_{M^{g}}^{LL}$ in Eq.(\ref{Rterms}) 
is enhanced by $\tan{\beta}$, due to the term proportional to $Y_b$.
For instance, for $\tan \beta \sim 10$, the value of 
$R^{LL}_{M^g}$ is reduced to
$R^{LL}_{M^g}\simeq -0.023$, while $R^{RL}_C$ is slightly increased to
$R^{RL}_C\simeq -0.033$ and the amplitudes ratio becomes
\be
R_A \simeq 
0.12 (\delta^u_{LL})_{31} +
 0.54(\delta^u_{LL})_{32} - 0.05 
(\delta^u_{RL})_{31}  - 0.21 (\delta^u_{RL})_{32}.
\ee
Furthermore it is remarkable to notice that, 
with heavy SUSY particles ($M_{\tilde{q}}\sim 1$ TeV), 
the $Z$-penguin diagram would provide the dominant contributions to 
$F_{\chi}$, since $R_C^{RL}$ tends to a constant value of order $-0.05$. 
This effect clearly shows the phenomena of non-decoupling of 
the chargino contribution to the Z penguin, as discussed for instance 
in Ref.\cite{Isidori}.

Finally, we stress that 
the contribution of $(\delta^u_{LL})_{32}$ 
to the chromomagnetic dipole operator, which leads to 
the dominant contribution to $S_{\phi K_S}$, is strongly 
constrained by $b\to s \gamma$ (which is particularly sensitive to
$C_{11}(\mu_W)$ ). This is due to the fact that 
$(\delta^u_{LL})_{32}$ gives almost the same contribution to both 
$C_{11}(\mu_W)$ and $C_{12}(\mu_W)$, as can be seen from Eq.(\ref{Rterms}).
Notice that this is not the case for gluino exchanges, since there
the contributions to the chromomagnetic dipole operator are enhanced by 
color factors with respect to the magnetic dipole ones,
allowing large contributions to $C_{12}$ while respecting the $b\to s\gamma$
constraints \cite{CGG}.
Regarding the effects of $(\delta^u_{RL})_{31}$ and $(\delta^u_{LL})_{31}$, 
their contributions to $S_{\phi K_S}$ is quite small since 
they are mostly constrained by $\Delta M_B$ and $\sin 2 \beta$ 
\cite{Gabrielli:2002fr}.

For the above set of input parameters, the $b\to s \gamma$ 
limits impose $\vert (\delta^u_{LL})_{32} \vert < 0.58$. Thus,  
the maximum individual mass insertion contributions are given by 
$ \left\vert \frac{A^{SUSY}_{LL32}}{A^{SM}}\right\vert < 0.31$ and
$ \left\vert \frac{A^{SUSY}_{RL32}}{A^{SM}}\right \vert < 0.21.$
This shows that after imposing the $b\to s \gamma$ constraints, 
the contribution from 
$(\delta^u_{LL})_{32}$ is of the same order as the contribution 
from $(\delta^u_{RL})_{32}$. 

Since the ratio $R_{\phi}\equiv |A^{SUSY}/A^{SM}| < 1$, 
one can expand the expression of
$S_{\phi K_S}$ in Eq. (\ref{cpmixing}) in terms of $R_{\phi}$ and gets the 
following simplified formula
\be
S_{\phi K_S} = \sin 2 \beta + 2 \cos 2 \beta~ \sin \theta_{\phi}~ R_{\phi},
\ee
which shows that with $R_{\phi} \sim 0.4$ and even if 
$\sin \theta_{\phi}\sim -1$,
one can reduce $S_{\phi K_S}$ from the SM prediction $\sin 2 \beta$ 
to $0.2$ at most and
it is not possible with one mass insertion contribution to reach 
negative CP asymmetry.
However, by considering the contributions from both
$(\delta^u_{LL})_{32}$ and $(\delta^u_{RL})_{32}$ simultaneously, 
$R_{\phi}$ can become large and values of order
$S_{\phi K_S}\simeq -0.2$ can be achieved.

It is worth mentioning that we have also considered the BR of $B^0\to \phi K^0$
decay and ensured that the SUSY effects do not violate the experimental limits 
observed by BaBar and Belle \cite{Br}.

Finally, let us emphasize that generally in supersymmetric models 
the lighter chargino is expected to be one of the lightest 
sparticles (for instance, in Anomaly Mediated SUSY breaking 
models it is almost degenerate with the lightest one).  
Thus, it can be expected to contribute significantly in the 
one-loop processes.
Although the gluino contribution to the studied asymmetry can
be very large, on the other hand gluino in many models is one
of the heaviest SUSY partners and thus its contribution may be
reduced essentially.

To conclude, we have studied the chargino contributions to the 
CP asymmetry $S_{\phi K_S}$ and
showed that, although the experimental limits on $b \to s \gamma$ 
impose stringent constraints
on the parameter space, it is still possible to reduce 
$S_{\phi K_S}$ significantly and negative values within the
$1\sigma$ experimental range can be obtained.

\section*{Acknowledgments} 
DC, EG, and KH thank the Academy
of Finland (project number 48787) for financial support.
SK would like to thank the Helsinki Institue of Physics for its 
kind hospitality.
\section*{Appendix}
Here we provide the analytical results for the 
the expressions $R_{F}$ and $\bar{R}_{F}$ appearing in Eq.(\ref{Fch}), 
which are given by

\bea
R_D^{LL}&=&\sum_{i=1,2} \,
|V_{i1}|^2 \, x_{\W i}\, P_D(x_i)
\nonumber
\\
R_D^{RL}&=&-\sum_{i=1,2} V_{i2}^{\star}V_{i1} \, x_{\W i}\, P_D(x_i)
\nonumber
\\
R_D^{RR}&=&\sum_{i=1,2} \, |V_{i2}|^2 \, x_{\W i}\, P_D(x_i)
\nonumber
\\
R_D^{LR}&=&\left(R_D^{RL}\right)^{\star}
\nonumber
\\
R_E^{LL}&=&\sum_{i=1,2} \,
|V_{i1}|^2 \, x_{\W i}\, P_E(x_i)
\nonumber
\\
R_E^{RL}&=&-\sum_{i=1,2} V_{i2}^{\star}V_{i1} \, x_{\W i}\, P_E(x_i)
\nonumber
\\
R_E^{RR}&=&\sum_{i=1,2} \, |V_{i2}|^2 \, x_{\W i}\, P_E(x_i)
\nonumber
\\
R_E^{LR}&=&\left(R_E^{RL}\right)^{\star}
\nonumber
\\
R_C^{LL}&=&
\sum_{i=1,2}|V_{i1}|^2 \, P_C^{(0)}(\bar x_i)
+\sum_{i,j=1,2} \left[ U_{i1}V_{i1}U_{j1}^{\star}V_{j1}^{\star}\, 
P_C^{(2)}(x_i,x_j) \right. 
\nonumber\\
&+& \left. |V_{i1}|^2 |V_{j1}|^2
\left(\frac{1}{8}-P_C^{(1)}(x_i, x_j)\right)\right]
\nonumber
\\
R_C^{RL}&=&-\frac{1}{2}
\sum_{i=1,2}\, V_{i2}^{\star}V_{i1}\,  P_C^{(0)}(\bar x_i)
- \sum_{i,j=1,2}\, V_{j2}^{\star}V_{i1}\left(
U_{i1}U_{j1}^{\star}\, P_C^{(2)}(x_i, x_j)
\right.\nonumber \\
&+&
\left. 
V_{i1}^{\star} V_{j1}\, P_C^{(1)}(x_i, x_j)\right)
\nonumber
\\
R_C^{LR}&=&\left(R_C^{RL}\right)^{\star},
\nonumber
\\
R_C^{RR}&=&
\sum_{i,j=1,2}\, V_{j2}^{\star}V_{i2}\left(
U_{i1}U_{j1}^{\star}\, P_C^{(2)}(x_i, x_j) + 
V_{i1}^{\star} V_{j1}\, P_C^{(1)}(x_i, x_j)\right)
\nonumber
\\
R_{B^{u}}^{LL}&=&
2\sum_{i,j=1,2}\, V_{i1} V_{j1}^{\star}
U_{i1}U_{j1}^{\star}\,x_{Wj}\sqrt{x_{ij}}~
P_{B}^{u}(\bar x_j, x_{ij})
\nonumber
\\
R_{B^{u}}^{RL}&=&-
2\sum_{i,j=1,2}\, V_{i1} V_{j2}^{\star}
U_{i1}U_{j1}^{\star}\,x_{Wj}\sqrt{x_{ij}}~
P_{B}^{u}(\bar x_j, x_{ij})
\nonumber
\\
R_{B^{u}}^{LR}&=&\left(R_{B^{u}}^{RL}\right)^{\star}
\nonumber
\\
R_{B^{u}}^{RR}&=&
2\sum_{i,j=1,2}\, V_{i2} V_{j2}^{\star}
U_{i1}U_{j1}^{\star}\,x_{Wj}\sqrt{x_{ij}}~
P_{B}^{u}(\bar x_j, x_{ij})
\nonumber
\\
R_{B^{d}}^{LL}&=&
\sum_{i,j=1,2}\, |V_{i1}|^2 |V_{j1}|^2 \,x_{Wj}\, 
P_{B}^{d}(\bar x_j, x_{ij})
\nonumber
\\
R_{B^{d}}^{RL}&=&-
\sum_{i,j=1,2}\, V_{i2}^{\star} V_{i1}
|V_{j1}|^2\,x_{Wj}\, P_{B}^{d}(\bar x_j, x_{ij})
\nonumber
\\
R_{B^{d}}^{LR}&=&\left(R_{B^{d}}^{RL}\right)^{\star}
\nonumber
\\
R_{B^{d}}^{RR}&=&
\sum_{i,j=1,2}\, V_{i2}^{\star} V_{i1}
V_{j1}^{\star}V_{j2}\, x_{Wj}\, P_{B}^{d}(\bar x_j, x_{ij})
\nonumber
\\
R_{M^{\gamma,g}}^{LL}&=&\sum_i |V_{i1}|^2\,
x_{Wi}\, P_{M^{\gamma,g}}^{LL}(x_i)
-Y_b\sum_i V_{i1} U_{i2}\,
x_{Wi}\, \frac{m_{\chi_i}}{m_b} P_{M_{\gamma,g}}^{LR}(x_i)
\nonumber
\\
R_{M^{\gamma,g}}^{LR}&=&-\sum_i V_{i1}^{\star}V_{i2}\,
x_{Wi}\, P_{M_{\gamma,g}}^{LL}(x_i)
+Y_b\sum_i  V_{i2} U_{i2}\,
x_{Wi}\, \frac{m_{\chi_i}}{m_b} P_{M_{\gamma,g}}^{LR}(x_i)
\nonumber
\\
R_{M^{\gamma,g}}^{RL}&=&
-\sum_i V_{i1}V_{i2}^{\star}\,
x_{Wi}\, P_{M_{\gamma,g}}^{LL}(x_i)
\nonumber
\\
R_{M^{\gamma,g}}^{RR}&=&
\sum_i |V_{i2}|^2\,
x_{Wi}\, P_{M_{\gamma,g}}^{LL}(x_i)
\label{Rfunctions}
\eea
where $x_{\W i}=m_W^2/m_{\chi_i}^2$, 
$x_{i}=m_{\chi_i}^2/\tilde{m}^2$, $\bar x_i =\tilde{m}^2/m_{\chi_i}^2$, and
$x_{ij}=m_{\chi_i}^2/m_{\chi_j}^2$.
The expressions for the functions $P_{E,D,C}$, $P_{B}^{(u,d)}$, 
$P_{M_{\gamma,g}}^{LL}$, and $P_{M_{(\gamma,g)}}^{LR}$ ,
are given in the next subsection.

There are other contributions which come from box diagrams, where both
chargino and gluino are exchanged ($B_{\tilde{g}}^{u,c}$), 
and cannot be expressed in the same form of Eq.(\ref{Fch}).  
We provide below the results for these
contributions, which affect only the Wilson coefficients 
$C_{1,2}^{(u,c)}(\mu_W)$ as
\bea
C^{(u,c)}_1(\mu_W)
&=&\frac{\alfasw}{16\pi}\left(14-B^{(u,c)}_{\tilde{g}}\right)\nonumber \\
C^{(u,c)}_2(\mu_W)&=& 1+\frac{\alfasw}{48\pi}B^{(u,c)}_{\tilde{g}}
\eea
where
\bea
B_{\tilde{g}}^{u}&=&
\Big[\sum_{a} K^{\star}_{a2} K_{13}\du{LL}{1a}
\Big] R_{\tilde{g}}^{LL}(u)+
\Big[\sum_{a}
K^{\star}_{12} K_{a3}\du{LL}{a1}\Big] 
\left(R_{\tilde{g}}^{LL}(u)\right)^{\star}
\nonumber \\
&+&
\Big[\sum_{a} K^{\star}_{1a} K_{13}\dd{LL}{a2}\Big] R_{\tilde{g}}^{LL}(d)
 +
\Big[\sum_{a}
K^{\star}_{12} K_{1a}\dd{LL}{3a}\Big] 
\left(R_{\tilde{g}}^{LL}(d)\right)^{\star}
\nonumber \\
&+&
\Big[\sum_{a} K^{\star}_{12} K_{33}\du{RL}{31}\Big]Y_t\, R_{\tilde{g}}^{RL}
\label{boxgluinoU}
\eea
\bea
B_{\tilde{g}}^{c}&=&
\Big[\sum_{a} K^{\star}_{a2} K_{23}\du{LL}{2a}\Big]
R_{\tilde{g}}^{LL}(u)+
\Big[\sum_{a} K^{\star}_{22} K_{a3}\du{LL}{a2} \Big] 
\left(R_{\tilde{g}}^{LL}(u)\right)^{\star}
\nonumber \\
&+&
\Big[\sum_{a} K^{\star}_{2a} K_{23}\dd{LL}{a2}\Big]R_{\tilde{g}}^{LL}(d)
 +
\Big[\sum_{a}
K^{\star}_{22} K_{2a}\dd{LL}{3a}\Big] 
\left(R_{\tilde{g}}^{LL}(d)\right)^{\star}
\nonumber \\
&+&
\Big[\sum_{a} K^{\star}_{22} K_{33}\du{RL}{32}\Big] Y_t\, 
R_{\tilde{g}}^{RL}
\label{boxgluinoC}
\eea
and the functions $R_i$ are given by
\bea
R_{\tilde{g}}^{LL}(u)&=&4 x_{W\tilde{g}} \, \sum_{i=1,2}\Big[
|V_{i1}|^2 P_B^{d}(z_i,y) +2 U_{i1}V_{i1}
\left(\frac{m_{\chi_i}}{m_{\tilde{g}}}\right) P_B^{u}(z_i,y)\Big]
\\
R_{\tilde{g}}^{LL}(d)&=&4x_{W\tilde{g}} \, \sum_{i=1,2}
\Big[|U_{i1}|^2 P_B^{d}(z_i,y) 
+2 U_{i1}^{\star}V_{i1}^{\star}
\left(\frac{m_{\chi_i}}{m_{\tilde{g}}}\right) P_B^{u}(z_i,y)\Big]
\\
R_{\tilde{g}}^{RL}&=&-4x_{W\tilde{g}} \, \sum_{i=1,2}
\Big[V_{i1}V_{i2}^{\star} P_B^{d}(z_i,y) 
+2 V_{i2}^{\star}U_{i1}^{\star}
\left(\frac{m_{\chi_i}}{m_{\tilde{g}}}\right) P_B^{u}(z_i,y)\Big]
\label{Rfunctgluino}
\eea
with $x_{W\tilde{g}}=m_W^2/m_{\tilde{g}}^2$,
$z_i=m_{\chi_i}^2/m_{\tilde{g}}^2$, and $y=\tilde{m}^2/m_{\tilde{g}}^2$.
In obtaining the above results in Eqs.(\ref{boxgluinoU})-(\ref{boxgluinoC})
we neglect terms of order of ${\cal O}(Y_b)$.

\subsection*{Loop functions}
Here we provide the expressions for the loop functions of penguin
$P_{D,E,C}$, box $P_B^{(u,d,\tilde{g})}$, and
magnetic-- and chromomagnetic--penguin diagrams
$P_{M_{\gamma,g}}^{LL}$, and $P_{M_{\gamma,g}}^{LR}$ 
respectively, which enter in Eqs.(\ref{Rfunctions}),(\ref{Rfunctgluino})
\bea
P_D(x)&=&
{\frac{2\,x\,\left( -22 + 60\,x - 45\,{x^2} + 4\,{x^3} + 
       3\,{x^4} - 3\,\left( 3 - 9\,{x^2} + 4\,{x^3} \right) \,
        \log{x} \right) }{27\,{{\left(1- x  \right) }^5}}}
\nonumber \\
P_E(x)&=&{\frac{x\,\left( -1 + 6\,x - 18\,{x^2} + 10\,{x^3} + 
         3\,{x^4} - 12\,{x^3}\,\log{x} \right) }{9\,
     {{\left(1- x \right) }^5}}}
\nonumber \\
P_C^{(0)}(x)&=&
{\frac{x\,\left( 3 - 4\,x + {x^2} + 2\,\log{x} \right) }
   {8\,{{\left(1- x  \right) }^3}}}
\nonumber \\
P_C^{(1)}(x,y)&=&\frac{1}{8\left(x-y\right)}\left[
\frac{x^2\left(x-1-\log{x}\right)}{(x-1)^2}
-\frac{y^2\left(y-1-\log{y}\right)}{(y-1)^2}\right]
\nonumber \\
P_C^{(2)}(x,y)&=&\frac{\sqrt{xy}}{4\left(x-y\right)}\left[
\frac{x\left(x-1-\log{x}\right)}{(x-1)^2}
-\frac{y\left(y-1-\log{y}\right)}{(y-1)^2}\right]
\nonumber \\
P_B^u(x,y)&=&
{\frac{-y - x\,\left( 1 - 3\,x + y \right) }
    {4\,{{\left( x -1  \right) }^2}\,
      {{\left( x - y \right) }^2}}} - 
  {\frac{x\,\left( {x^3} + y - 3\,x\,y + {y^2} \right) \,
      \log{x}}{2\,{{\left(x -1  \right) }^3}\,
      {{\left( x - y \right) }^3}}} 
\nonumber \\
&+&  {\frac{x\,y\,\log{y}}
    {2\,{{\left( x - y \right) }^3}\,\left(y -1  \right) }}
\nonumber \\
P_B^d(x,y)&=&-
{\frac{x\,\left( 3\,y - x\,\left( 1 + x + y \right)  \right) }
    {4\,{{\left(x -1  \right) }^2}\,
      {{\left( x - y \right) }^2}}} - 
  {\frac{x\,\left( {x^3} + \left( x -3 \right) \,{x^2}\,y + 
        {y^2} \right) \,\log{x}}{2\,
      {{\left(x -1  \right) }^3}\,{{\left( x - y \right) }^3}}
    } 
\nonumber \\
&+& {\frac{x\,{y^2}\,\log{y}}
    {2\,{{\left( x - y \right) }^3}\,\left(y -1 \right) }}
\nonumber \\
P_{M_{\gamma}}^{LL}(x)&=&- x \frac{d}{d x} \Big( 
x F_{1}(x)+\frac{2}{3}x F_2(x)\Big)
\nonumber \\
P_{M_{\gamma}}^{LR}(x)&=&- x \frac{d}{d x} \Big( 
x F_{3}(x)+\frac{2}{3}x F_4(x)\Big)
\nonumber \\
P_{M_{g}}^{LL}(x)&=&- x \frac{d}{d x} \Big(x F_2(x)\Big)
\nonumber \\
P_{M_{g}}^{LR}(x)&=&- x \frac{d}{d x} \Big(x F_4(x)\Big)
\eea
where the functions $F_i(x)$ are provided in Ref.\cite{BBMR}.
\subsection*{Light right-stop}
Here we generalize the above formulas for the case in which
the right-stop is lighter than other squarks.
Notice, 
that this will modify only the expressions of $R_F^{RL}$ and $R_F^{RR}$,
since the light right-stop  does not affect $R_F^{LL}$.  
In the case of $R_F^{RR}$ 
the functional forms of $R_F^{RR}$ remain unchanged, while the
arguments of the functions involved are changed as 
$x_i   \to   x_{it}$  and ${\bar x}_i \to {\bar x}_{it}$.
In the case of $R_F^{LR}$ and $R_F^{RL}$ 
the analytical expression of loop functions of penguin
$P_{D,E,C}$, box $P_B^{(u,d,\tilde{g})}$, and
magnetic and chromomagnetic penguin diagrams
$P_{M_{\gamma,g}}^{LL}$ and $P_{M_{\gamma,g}}^{LR}$ respectively,
should be changed as follows
\bea
P_D(x_i,x_{it})&=&{2\over{(x_t -1)}} 
\left[x_{it} D_\chi(x_{it}) - x_i D_\chi(x_i)\right]
\nonumber \\
P_E(x_i,x_{it})&=&{2\over{(x_t -1)}} 
\left[x_{it} E_\chi(x_{it}) - x_i E_\chi(x_i)\right]
\nonumber \\
P_C^{(1,2)}(x_i,x_{it}, x_j,x_{jt})&=& {2\over{(x_t -1)}}
  \left[ C_{\chi}^{(1,2)}(x_{jt},x_{it}) - C_{\chi}^{(1,2)}(x_{j},x_{i})\right]
\nonumber \\
P_C^{(0)}({\bar x}_i,{\bar x}_{it})&=&{4\over{(x_t -1)}}
  \left[ C_{\chi}^{(1)}({\bar x}_{it},{\bar x}_{i}) - 
C_{\chi}^{(1)}({\bar x}_{i},{\bar x}_{i}) \right]
\nonumber \\
P_B^{(u)}({\bar x}_j,{\bar x}_{jt}, x_{ij})&=&{1\over{2(x_t -1)}} 
\left[ B_{\chi}^{(u)}({\bar x}_{jt},{\bar x}_j, x_{ij})
-B_{\chi}^{(u)}({\bar x}_{j},{\bar x}_j, x_{ij}) \right]
\nonumber \\
P_B^{(d)}({\bar x}_j,{\bar x}_{jt}, x_{ij})&=&-{1\over{2(x_t -1)}} 
\left[B_{\chi}^{(d)}({\bar x}_{jt},{\bar x}_j, x_{ij})
-B_{\chi}^{(d)}({\bar x}_{j},{\bar x}_j, x_{ij}) \right]
\nonumber \\
P_{M_{\gamma}}^{LL}(x_i,x_{it})&=&{1\over{(x_t -1)}} 
\left[x_{it}\left(  F_{1}(x_{it})+\frac{2}{3} F_2(x_{it})\right)
-x_{i}\left(  F_{1}(x_{i})+\frac{2}{3} F_2(x_{i})\right)\right]
\nonumber \\
P_{M_{\gamma}}^{LR}(x_i,x_{it})&=&{1\over{(x_t -1)}} 
\left[x_{it}\left(  F_{3}(x_{it})+\frac{2}{3} F_4(x_{it})\right)
-x_{i}\left(F_{3}(x_{i})+\frac{2}{3} F_4(x_{i})\right)\right]
\nonumber \\
P_{M_{g}}^{LL}(x_i,x_{it})&=&{1\over{(x_t -1)}} 
\left[x_{it} F_2(x_{it}) -x_{i} F_2(x_{i})\right]
\nonumber \\
P_{M_{g}}^{LR}(x_i,x_{it})&=&{1\over{(x_t -1)}} 
\left[x_{it} F_4(x_{it}) -x_{i} F_4(x_{i})\right]
\label{Frightstop}
\eea
where  $x_{i}=m_{\chi_i}^2/\tilde{m}^2$, $
{\bar x_i} =\tilde{m}^2/m_{\chi_i}^2$, 
$x_{it}=m_{\chi_i}^2/{m^2_{{\tilde t}_R}}$,
 ${\bar x_{it}} ={m^2_{{\tilde t}_R}}/m_{\chi_i}^2$, 
$x_{ij}=m_{\chi_i}^2/m_{\chi_j}^2$ and $x_t={m^2_{{\tilde t}_R}}/\tilde{m}^2$.
The functions $D_{\chi},C_{\chi},E_{\chi},C^{(1,2)}_{\chi},
B^{(u,d)}_{\chi}$ and $F_i$ are provided in Ref.\cite{GG} and Ref.\cite{BBMR}
respectively.

%

\vspace{1cm}

\end{document}